\documentclass[a4paper]{jpconf}
\usepackage{graphicx}
\begin{document}

\title{Stability of charge-stripe ordered La$_ {2-x}$Sr$_{x}$NiO$_{4+\delta}$ at one third doping.}

\author{P. G. Freeman$^{1*}$, R. A. Mole$^{2}$, N. B. Christensen$^{3,4}$, A. Stunault$^{5}$, D. Prabhakaran$^6$}

\address{$^1$Jeremiah Horrocks Institute for Mathematics, Physics, and Astronomy, University of Central Lancashire, Preston, PR1 2HE, UK.}

\address{$^2$Heinz Maier Leibnitz Zentrum (MLZ), Teschnische Universitaet Muenchen Lichtenbergstrasse 1, Garching 85747, Germany.}

\address{$^3$ Laboratory for Neutron Scattering, ETHZ and PSI,
CH-5232 Villigen PSI, Switzerland. }

\address{$^4$ Department of Physics, Technical University of Denmark (DTU), 2800 Konigs Lyngby, Denmark. }

\address{$^5$Institut Laue-Langevin, BP 156, 38042 Grenoble Cedex 9, France.}

\address{$^6$Department of Physics, Oxford University, Oxford, OX1 3PU, United Kingdom.}

\ead{pgfreeman@uclan.ac.uk}

\pagenumbering{roman}

\begin{abstract}

The stability of charge ordered phases is doping dependent, with different materials having particularly stable ordered phases. In the half filled charge ordered phases of the cuprates this occurs at one eighth doping, whereas in charge-stripe ordered La$_ {2-x}$Sr$_{x}$NiO$_{4+\delta}$ there is enhanced stability at one third doping. In this paper we discuss the known details of the charge-stripe order in La$_ {2-x}$Sr$_{x}$NiO$_{4+\delta}$, and how these properties lead to the one third doping stability.\\


\end{abstract}

\section{Introduction}

Since the discovery of charge-stripes in La-based cuprates after their theoretical prediction,  the role of charge order in the cuprates has been under great debate\cite{tranquada-Nature-1995,Bianconi-PRL-1996,Zaaneen-PRB-1989,Poilblanc-PRB-1989}.  Recently it has been discovered that charge order is not restricted to La based cuprates, but found in many different families of cuprates near one eighth doping\cite{Comin}. With the question of whether or not there is a common dimensionality of the charge order in these different cuprate materials currently under discussion\cite{Comin}.  Is the charge order from one dimensional charge-stripes or two dimensional bi-directional checkerboard charge order? In the La-based cuprates the charge order appears to be stabilized at one eighth doping, with a saturation of the wavevector of the charge and spin super-structures, as well as highest charge ordering temperatures\cite{Comin,Yamadaplot}.  To better understand the role of charge order in the cuprates, charge order in non-superconducting materials  have been studied in detail.

\begin{figure}[!h]
\begin{center}
\includegraphics[clip=,width = 11 cm]{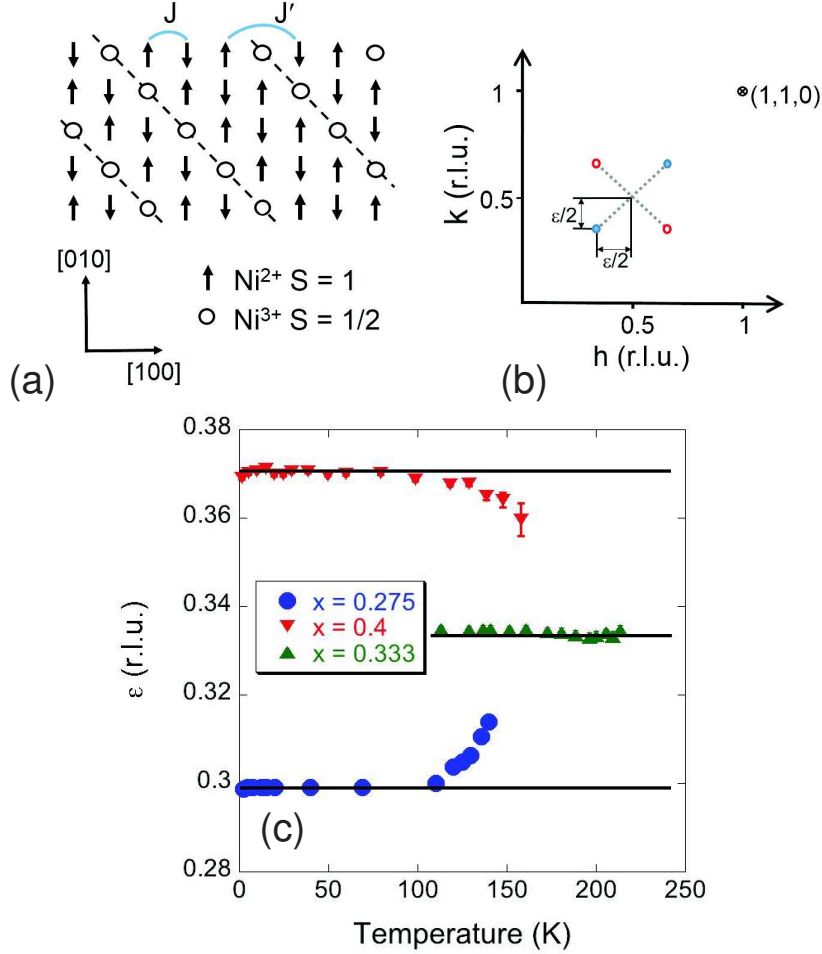}
\caption[Tempseepofincommenusrability ]{(a) A Ni-centred cartoon model of charge-stripe order in La$_ {2-x}$Sr$_ {x}$NiO$_ {4+\delta}$ (LSNO) $x \sim 0.26-0.29$ showing the Ni sites of a single Ni-O plane, for a doping level below one third doping that leads to incommensurate order. We indicate the  intra-stripe spin interaction J, and inter-stripe spin interaction  $J'$ that are necessary to describe the magnetic excitation spectrum from the ordered Ni$^ {2+}$ S = 1 spins of LSNO\cite{boothroyd-PRB-2003,Woo}.  (b) A diagram of first Brillouin zone of the $(h, k, 0)$ plane in reciprocal space for LSNO. The location of magnetic Bragg reflections of the two spin stripe domains are represented by circles, with the filled blue circles representing spin-stripes as orientated in (a), and the $(1, 1, 0)$  Bragg reflection is indicated. Charge-stripe order Bragg reflections would similarly be observed at $(h \pm \varepsilon, k \pm \varepsilon, l)$, for $l = $ odd positions in reciprocal space. The average periodicity of the charge-stripe order is given by $P = 1/\varepsilon$. (c)The temperature evolution of the incommensurability, $\varepsilon$ of charge-stripe ordered La$_ {2-x}$Sr$_{x}$NiO$_{4}$ $x = 0.275, 0.333, 0.4$ obtained by neutron diffraction\cite{freeman-PRB-2011,freeman-PRB-2004,prab}. On warming towards the spin ordering temperature ( $T_ {SO}(x = 0.275)  = 130\pm 10$\, K, $T_ {SO}(x = 0.4)  = 150\pm 10$\, K) the value of $\varepsilon$ tends towards $1/3$\cite{kajimoto-PRB-2001,Ishizaka-PRL-2004}. The tendency of $\varepsilon \rightarrow 1/3$ is believed to be the result of the charge ordering preference\cite{kajimoto-PRB-2001,Ishizaka-PRL-2004}.} \label{fig:elastic}
\end{center} \end{figure}

The insulating La$_ {2-x}$Sr$_{x}$NiO$_{4+\delta}$ (LSNO) is a model system to study charge-stripe order, with static charge-stripes whose correlation lengths  exceed 300\,\AA\ near one third doping\cite{neutron,Tranquada-PRL-1994,Sachan-PRB-1995,Tranquada-PRB-1996,x-ray,chen93,Yamada94,pash-PRL-2000,Yoshinari-PRL-2000,Ulbrich-PhysicaC-2002}. This material has a body centred tetragonal structure that is pseudo-issostructural with `214' cuprates. There are other materials such as La$_ {2-x}$Sr$_{x}$CoO$_{4+\delta}$ which charge order, but La$_ {2-x}$Sr$_{x}$CoO$_{4+\delta}$   $x \sim 1/3$ has both charge-stripe and checkerboard charge order\cite{dress-NatComm-2013,Babkevich_NatComm_2016}, whereas LSNO for $0.135 \leq n_{h} < 0.5$ only has charge-stripe order\cite{yoshizawa-PRB-2000,hatton-2002}. 
Charge-stripe order in LSNO  occurs in the square lattice NiO$_2$ layers, consisting of diagonal bands of antiferromagnetically (AFM) ordered Ni$^{2+}$ $S = 1$ spins separated by charged domain walls that act as antiphase boundaries to the magnetic order. On cooling the charge-stripes first order, before the spin stripe magnetic order occurs at a lower temperature\cite{yoshizawa-PRB-2000}. In diffraction techniques such as x-ray diffraction this produces superstructure charge-stripe order Bragg reflections at $(h \pm \varepsilon, k \pm \varepsilon, l)$, for $l = $ odd  positions in reciprocal space and $\varepsilon \approx n_h$, where $n_h = x +2\delta$ is the hole doping. The periodicity of the charge-stripe order is therefore given by $P = 1/\varepsilon$, and $\varepsilon$ is known as the incommensurability. Whereas the magnetic Bragg reflections are observed in neutron diffraction at $(h + 1/2 \pm \varepsilon /2, k + 1/2 \pm \varepsilon /2, l)$ positions in reciprocal space. As LSNO has a tetragonal symmetry the charge-stripes run along both diagonals of the NiO$_2$ square lattice, with an equal population of the two twins\cite{boothroyd-PRB-2003,Woo}. The spins of the charge-stripe electrons themselves do not magnetically order, but are anitferromagnetically correlated over a short distance along the charge-stripes\cite{boothroyd-PRL-2003,freeman-PRB-2011}.

In La$_ {2-x}$Sr$_{x}$NiO$_{4+\delta}$ (LSNO) charge-stripe order at one third doping is observed to be particularly stable\cite{yoshizawa-PRB-2000,ramirez-PRL-1996,lee-PRB-2001,kajimoto-PRB-2001,freeman-PRB-2004,Ishizaka-PRL-2004}. For instance the charge-stripe and spin-stripe transition temperatures have a maximum at one third doping\cite{yoshizawa-PRB-2000}, with a maximum  in transition temperature for a spin reorientation of the ordered moments at one third doping\cite{lee-PRB-2001,freeman-PRB-2004,Giblin-PRB-2008}. The charge ordering at one third doping causes a  sound velocity hardening not observed at other doping levels, with specific heat at one third doping observing an entropy change consistent with charge ordering\cite{ramirez-PRL-1996}. Diffraction measurements reveal a systematic doping variation of $\varepsilon$ away from the expected value of  $n_h = \varepsilon $ for charge-stripes with 1 hole per lattice site, which we will discuss in detail later in this report. In figure \ref{fig:elastic} we show (a) a toy model of charge-stripe order in a Ni-O plane, (b) the positions of magnetic Bragg reflections from charge-stripe order, and
(c) the striking temperature evolution of $\varepsilon$ for three different doping levels of LSNO. The samples in Fig. \ref{fig:elastic} (c) are $x = 0.275$, $x = 0.333$ and $x = 0.4$, showing data that was taken in previous studies\cite{freeman-PRB-2011,freeman-PRB-2004,prab}.  From neutron diffraction Kajimoto $\it {et. al.}$ and from x-ray diffraction Ishizaka  $\it {et. al.}$  observed for charge-stripe ordered in LSNO with $n_h \neq 1/3$, on increasing temperature towards the ordering temperature the $\varepsilon$ changes and tends towards a value of 1/3\cite{kajimoto-PRB-2001,Ishizaka-PRL-2004}. The purpose of this paper is to highlight key points of our understanding of charge-stripe order in LSNO, particularly to elucidate why charge-stripe order is particularly stable at one third doping. This paper therefore builds and updates a previous  review of neutron scattering studies on LSNO\cite{Ulbrich-PhysicaC-2002}.

\section{Theoreticl Studies}

\subsection{Charge order formation.}

Theoretical calculations for the ground state of holes doped on to a two dimensional metal oxide square lattice first predicted the formation of charge-stripes before their experimental observation\cite{Zaaneen-PRB-1989,Poilblanc-PRB-1989}.  Different approaches were undertaken, all of which showed that electron-electron interactions and electron-phonon interactions help to stablize charge-stripe order\cite{Poilblanc-PRB-1989,Zaaneen-PRB-1994}. Specifically the interactions result in electron hopping and the Coulomb interaction leads to charge-stripe formation\cite{Poilblanc-PRB-1989,Zaaneen-PRB-1994, Raczkowski-PRB-2006,Schwing}. These models show that for charge-stripe order in LSNO there is an occupancy equivalent to one hole per charge-stripe site, but this hole has an extended size within the plane\cite{Zaaneen-PRB-1989,Poilblanc-PRB-1989,Zaaneen-PRB-1994}.  Recent modelling of the charge-stripe order in LSNO has also shown the potential importance of the Jahn-Teller effect in charge-stripe formation\cite{Rosciszewski}.

The successful prediction of charge-stripe order before it's discovery shows the importance of theoretical work for our understanding charge-stripe order LSNO, however there are limitations to these models. The theoretical models do not include temperature, as discussed in section one the charge-stripe order has a complicated temperature dependence, highlighting the need to develop theoretical models to account for this behaviour. Theoretical papers have invariably predicted a ferromagnetic spin interaction between the charge-stripe electrons\cite{Zaaneen-PRB-1994}, when this spin interactions is observed by inelastic neutron scattering to be antiferromagnetic\cite{boothroyd-PRL-2003,freeman-PRB-2011}. Antiferromagnetic interactions between the charge-stripe electrons have been theoretically predicted  along with orbital ordering when lattice distortions are considered, but we are unaware of any observation for orbital ordering in LNSO\cite{Hotta-PRL-2004}.

\section{Experimental Studies}

\subsection{Centring and physical extent of charge-stripes.}

Different experimental techniques have been deployed to study the centring of charges-stripes in LSNO. Neutron diffraction of oxygen doped La$_ {2}$NiO$_{4+\delta}$, $\delta = 2/15$, studied the dependence of the magnetic Bragg reflections along the $(0, 0, l)$ direction out of the Ni-O planes, determining that below the spin ordering temperature the charge-stripes are predominantly Ni centred over oxygen centred\cite{wochner-PRB-1998}. This study also observed higher order harmonic magnetic Bragg reflections, with the observed ratio of the first to third harmonic magnetic Bragg reflections being consistent with Ni centred charge-stripes\cite{PHILLABAUM}. A later study of LSNO $x = 0.275$ however observed no third harmonic magnetic Bragg reflection\cite{freeman-JPCS-2012}, possibly indicating the stabilization of charge-stripe order by a lattice distortion connected to the oxygen super cell structure in La$_ {2}$NiO$_{4+\delta}$, $\delta = 2/15$\cite{wochner-PRB-1998}.

Tunnelling electron microscope (TEM) on the surface of LSNO $x = 0.275$ determined the charge-stripes to have two lattice positions, Ni or O centred, with predominantly Ni centred charge-stripes\cite{li-PRB-2003}. This study confirmed the finding of diffraction studies  that charge-stripes run parallel to each other in neighbouring Ni-O planes, unlike Nd doped La-based cuprates\cite{tranquada-Nature-1995}. Charge-stripes are observed by TEM to have an extended nature of at least 1 Ni-Ni spacing perpendicular to the charge-stripe direction within the Ni-O plane. The energy and polarization dependence observed in resonant soft x-ray diffraction studies of charge-stripe order several hundred \AA\  into the bulk of  LSNO  $x = 0.2$ can be explained within a Ni centred charge-stripe model with the hole located predominantly on the oxygens surrounding the Ni\cite{schu-PRL-2005}. This can be compared to NMR observation of Ni centred charge-stripes consistent with $S =  1/2$ Ni$^ {3+}$ 3d$^ {7}$ ions\cite{NMR}, the oxygen character of holes observed by X-ray Absorption spectrocopy\cite{XAS}, and the determination of the small polaron size of the hole from optical spectroscopy\cite{polaron}. All of these findings consistently indicate charge-stripes in La$_ {2-x}$Sr$_ x$NiO$_{4+\delta}$ can be Ni or O centred with Ni centring dominating at low temperature. The charge-stripes extend more than one lattice spacing perpendicular to the charge-stripe direction within the Ni-O plane, and the holes have a greater fraction on the O sites rather than central Ni site. 

The experimental determination of a preference for Ni centred charge-stripes and charge-stripe holes with an extended in-plane size, is consistent with the earlier theoretical predictions\cite{Poilblanc-PRB-1989,Zaaneen-PRB-1994}. The presence of both O and Ni centred charge-stripes indicates the close proximity of the two types of charge-stripes in LSNO, potentially responsible for low 5-7\,meV gapped collective dynamics observed in optical conductivity measurements\cite{lloyd-hughes-PRB-2008}.

\subsection{Out of Ni-O plane Charge-stripe Stacking}

In Nd doped La-based cuprates the charge-stripe order out of the Cu-O plane rotates by 90 degrees between Cu-O layers due to pinning to the low temperature tetragonal structure of the lattice\cite{tranquada-Nature-1995}, with only a correlation between the charge-stripe order of the next nearest neighbour Cu-O planes  observed\cite{tranquadaaxxe-PRB-1996}. A significant difference of charge-stripe order in LSNO is that the charge-stripes are parallel to each other in neighbouring Ni-O planes\cite{li-PRB-2003}, and the $\approx 25$\, \AA\ out of Ni-O plane charge-stripe correlation length near one third doping indicates the charge order in 5-6 Ni-O planes are correlated\cite{yoshizawa-PRB-2000}. The charge-stripes of LSNO  are aligned parallel to each other  in adjacent layers to minimize the inter-layer Coulomb repulsion between charge-stripes\cite{wochner-PRB-1998}. The crystal structure of LSNO for $x >0.2$ is high temperature tetragonal (HTT), space group I4/mmm\cite{Sachan-PRB-1995}. In purely  oxygen doped LSNO  (i.e. $x = 0$, LNO) the oxygen orders in a supercell structure  which is single phase beyond  $\delta \approx 0.105$, and the supercell distorts the overall tetragonal structure\cite{tranquada-PRB-1994}. The HTT structure is body centred tetragonal with random tilting of Ni-O octhedra required to accommodate bond length mismatches. Along the crystal $(1, 1, 0)$ direction the Ni sites of adjacent layers are offset by half a lattice spacing. 

\begin{figure}[!h]
\begin{center}
\includegraphics[clip=, width = 15cm]{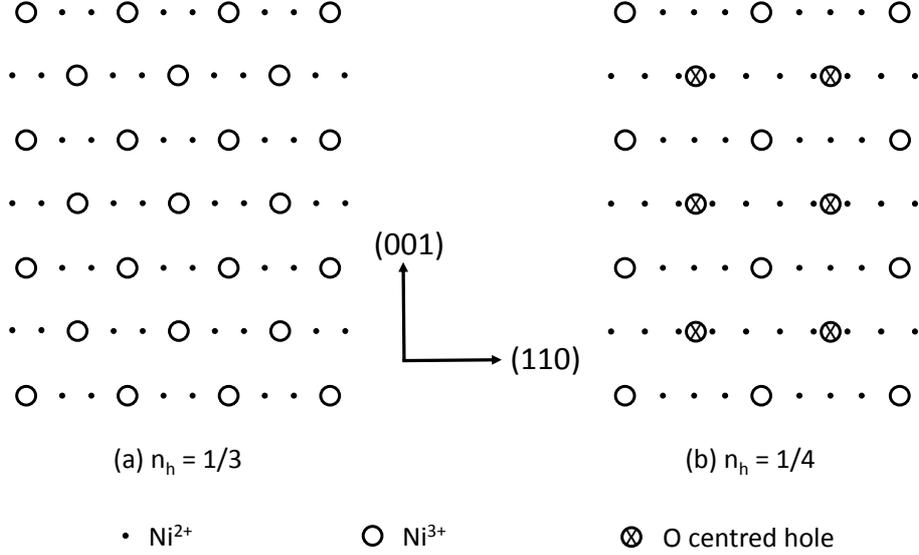}
\caption[Modeloflstacking ]{Stacking of charge-stripes along the c-axis in LSNO (a) Ni centred charge-stripes at one third doping, (b) alternating Ni and O centred charge-stripes at one quarter doping, where the charge-stripes are directed perpendicular to the figure. For clarity purposes O sites directly involved in the charge order and other atoms are omitted. Although there is only a preference for Ni centred charge-stripes so that O centred charge-stripes are observed, the occurrence of O centred charge-stripes appears uncorrelated\cite{wochner-PRB-1998,li-PRB-2003}, unlike the model in (b).
} \label{lstructure}.
\end{center} \end{figure}

Consider LSNO with only Ni centred charge-stripes and Coulomb repulsion between charge-stripes in different layers. For charge-stripes that are an odd number of Ni-spacings apart (3, 5, etc) the charge-stripes in one layer can have an equidistant separation from the charge-stripes in adjacent layers forming an ideal  body centred stacking structure, see figure \ref{lstructure} (a). If the charge-stripes are an even number of Ni-spacings apart, then the charge-stripes in one layer cannot have an equidistant separation from the charge-stripes in adjacent layers, e.g. for charge-stripes spaced 4 Ni-Ni spacings apart the charge-stripes in an adjacent layer would be offset by 1.5 or 2.5 Ni spacings creating electrostatic potential differences. In this way in LSNO for Ni centred charge-stripes and a charge-stripe spacing commensurate with the crystal structure, the Coulomb repulsion out of the Ni-O plane is minimised for charge-stripes an odd number of Ni-spacings apart. 

In the last section we discussed how in LSNO the charge-stripes are only preferentially centred on Ni sites, and  charge-stripes can also be centred on O sites\cite{wochner-PRB-1998,li-PRB-2003}.  If at $1/4$ doping in LSNO the charge-stripes were centred in equal amount on Ni or O sites then the adjacent layers could alternate between O and Ni centred stripes to maintain an offset of 2 Ni-Ni in-plane spacings between layers, as we show in figure \ref{lstructure} (b). But charge-stripes in LSNO are mainly Ni centred, and TEM of LSNO $x = 0.275$ indicated no preference for layers to have Ni or O centred charge-stripes\cite{li-PRB-2003}.  The co-operative switching between layers of O and Ni centred charge-stripes required for minimization of the inter-layer Coulomb repulsion at $1/4$ doping is therefore not realised, and we are unaware of any charge-stripe order stability at 1/4 doping in LSNO.


For LSNO doping that is not a simple fraction such as $1/3$ or $1/4$  etc. the material is thought of as being incommensurately doped, and the charge-stripes cannot be equidistantly offset between the adjacent Ni-O layers. Within a Ni-O layer for charge order that is either Ni or O centred, to achieve the average charge-stripe spacing in incommensurately doped materials, the spacing between charge-stripe must vary\cite{yoshizawa-PRB-2000}.  At one third doping for charge stripes that are orientated perpendicular to the $(1, 1, 0)$ tetragonal direction neutron diffraction see magnetic Bragg peaks at $(h + 1/2 \pm \varepsilon /2, k + 1/2 \pm \varepsilon /2, l)$, with $\varepsilon = 1/3$ and  $l = odd\ integer$ are observed as the charge-stripe order stacks in a body centred tetragonal structure. If  $\varepsilon \neq 1/3$ then neutron diffraction of stripe domains of the same orientation observes magnetic Bragg reflections at $(h + 1/2 \pm \varepsilon /2, k + 1/2 \pm \varepsilon /2, l)$, $l = odd$ or $even\ integer$, and the correlation length for the $l = even$ peaks is half that of the $l = odd$ peaks\cite{wochner-PRB-1998}. The $l = even$ peaks originate from non-ideal body centred stacking of the charge-stripes along $(0, 0, l)$\cite{freeman-PRB-2004}. Along with neutron diffraction, TEM showed the periodicity in incommensurately doped charge-stripe ordered LSNO varies randomly between layers\cite{wochner-PRB-1998,li-PRB-2003}. In a Ni-O plane where the charge-stripes space 3 Ni-Ni spacings for minimisation of the out-of-plane Coulomb interaction there is doping concentration in excess of chemical doping, leading to an increase in the in-plane Coulomb repulsion to a neighbouring in plane charge-stripe, so that a charge-stripe 4 Ni-Ni spacing apart relieves the local charge build up at the cost of the out-of-plane Coulomb interaction.  The Coulomb interaction between layers in incommensurately doped LSNO is therefore not fully minimized, and charge-stripe order in incommensurately doped LSNO is  less stable than commensurately doped LSNO $ x = 1/3$.


\subsection{Self Doping and Excess Doping}

Away from one third doping the $\varepsilon$ of the charge-stripe order systematically varies from the relationship $\varepsilon = n_{h}$ expected for charge-stripes with 1 hole per charge-stripe lattice site\cite{yoshizawa-PRB-2000}.  In LSNO the $\varepsilon$ is closer to one third than the chemical doping level, i.e. above one third doping level $\varepsilon <  n_{h}$\cite{yoshizawa-PRB-2000}.  For charge-stripes with 1 hole per lattice site, the holes in charge-stripes are localized, and insulating, so that the holes are observed to have a small polaron character\cite{polaron}. Below one third doping to achieve an $\varepsilon$ larger than expected from chemical doping, the material self-dopes by exciting electrons into a conduction band to increase the hole concentration in the valence band\cite{Katsufuji-PRB-1999}. The electrical properties of charge-stripe ordered in LSNO below one third doping are therefore from the promoted electrons, resulting in a negative Hall coefficient\cite{Katsufuji-PRB-1999}. Above one third doping, the holes required for the observed $\varepsilon$ indicates there is an excess of holes from chemical doping that do not take part in the charge-stripes, and electrical properties of the material are hole like, with a positive Hall coefficient\cite{Katsufuji-PRB-1999}. 

Below one third doping, increasing the doping level by self doping would reduce the average spacing between charge-stripes in the Ni-O plane. As the distance between charge-stripes in a single plane is significantly larger than the distance between charge-stripes of adjacent layers,
without a significant difference in the in- to out -of -plane electron screening the inter-plane charge-stripe repulsion is dominant. The reduction of out of Ni-O plane Coulomb interaction by ideal body centre stacking of  commensurate 3 Ni-Ni spacings must be energetically favourable over the cost increase of the Coulomb interaction within the Ni-O plane, as well as the energy cost of self doping. 
On increasing temperature  more electrons are thermally excited into the conduction states\cite{Katsufuji-PRB-1999}, further increasing the hole doping in the valence band. At the same time the electron hopping should become thermally active, consistent with the observed closing of a low energy optical gap in LSNO $ x = 0.275$\cite{lloyd-hughes-PRB-2008}, and  the reduced preference for Ni centred charge-stripes observed in in La$_ {2}$NiO$_{4+\delta}$, $\delta = 2/15$\cite{wochner-PRB-1998}. An increase in hole doping can therefore add additional charge-stripes to the order that can be accommodated into the ordered structure by the charge-stripes flexing apart, thus reducing the average periodicity towards 3 Ni-Ni spacings of the third doping. This explains the temperature dependence of $\varepsilon$ below one third doping\cite{kajimoto-PRB-2001,Ishizaka-PRL-2004}.

Above one third doping the cost of exclusion of holes from the charge order would be offset by several factors.  Near a third doping again the Coulomb interaction between planes would be reduced by being closer to the third doping's ideal body centre charge-stripe structure, offsetting the energy cost of exclusion of holes from the charge-stripe order\cite{Poilblanc-PRB-1989}. As the doping level increases towards half doping however the charge order loses its out of Ni-O plane correlation length indicating a loss of out-of plane Coulomb repulsion\cite{yoshizawa-PRB-2000,kajimoto-PRB-2003,Giblin-PRB-2008}. This is possibly due to the extended nature of the charge-stripes within the Ni-O plane leading to a poor out of plane definition of the charge-stripes\cite{li-PRB-2003,schu-PRL-2005}. At these higher doping levels a value of $\varepsilon$ nearer one third would reduce the Coulomb interaction between charge-stripes in the Ni-O planes by  increasing the charge-stripe spacing, and the spin order would be enhanced\cite{freeman-PRB-2005}, as we will discuss in the next section. The energy gain from the in plane Coulomb interaction and spin interactions would need to offset the energy cost of excluding a fraction of the holes form the charge-stripe order.  

As the temperature increases there is potential for further self doping of holes in the valence band driven by thermal excitation of the electrons into conduction the states\cite{Katsufuji-PRB-1999}, creating a larger excess of holes for doping levels above one third doping. Minimisation of the Coulomb interaction between planes by $\varepsilon$ tending towards one third would further exclude holes from charge-stripe with 1 hole per Ni site, with the hole chemical potential for excluded holes increasing the great the density of excluded holes is. The energy cost of excluding holes would therefore reduce the energy savings from minimisation of Coulomb and spin interactions. This would explain why  the  $\varepsilon$'s tendency to change towards one third on increasing temperature is smaller for doping levels above one third doping, than for doping levels below one third doping\cite{kajimoto-PRB-2001,Ishizaka-PRL-2004}, see Fig. \ref{fig:elastic} (c).

\subsection{Spin interactions, and their role on charge ordering near half doping}

Two spin interactions are responsible for the magnetism of the ordered Ni$^{2+}$ $S = 1$ moments in LSNO, $J$ a intra-spin stripe interaction between nearest neighbouring Ni sites of the spin stripes along the Ni-O-Ni bonds, and $J' $ an inter-spin stripe next nearest neighbour interaction across the charge-stripes that is parallel to the Ni-O bond direction\cite{boothroyd-PRB-2003,Woo,freeman-PRB-2005}, see Fig. \ref{fig:elastic} (A). Both the $J$ and $J'$ spin interactions have no significant doping dependence over the doping range $0 \leq n_{h} \leq 0.5$, with $J = 28$\,meV and $J' = 14$\,meV \cite{boothroyd-PRB-2003,Woo,freeman-PRB-2005,Nakajima-JPSJ-1993,freeman-PRB-2009}. The value of the equivalent nearest neighbour spin interaction in La$_2$CuO$_4$ is 136\,meV\cite{Hayden-LCO}. From the magnetic excitation spectrum of LSNO there is no observable spin interaction out of the Ni-O planes\cite{boothroyd-PRB-2003,Nakajima-JPSJ-1993}. This is consistent with the structure of the magnetic order arising from the charge-stripe structure\cite{wochner-PRB-1998}. Although there is no out-plane spin interaction, as the spin order develops there is a  reduction in out-of-plane correlation lengths of the spin order observed  in LSNO $x = 0.275$ and the charge order in LSNO $x = 1/3$, the latter being described by Landau theory as an order-disorder transition\cite{freeman-JPCS-2010,LeeHatton}.

\begin{figure}[!hb]
\begin{center}
\includegraphics[clip=, width = 8cm]{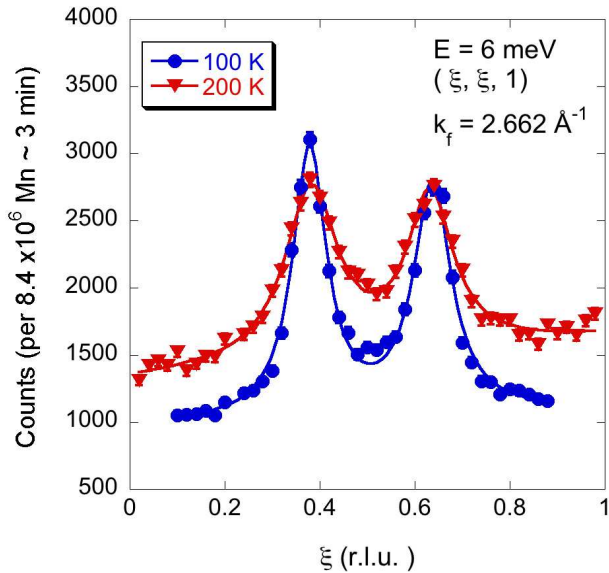}
\caption[Temperature dependence of magnetic excitations in La2NiO4.11]{Temperature dependence of magnetic excitations in charge-stripe ordered La$_2$NiO$_ {4.11}$\cite{details},  which magnetically orders at  $127.5 \pm 2.5$\,K and charge orders at $\sim 150$\, K\cite{Tranquada-PRB-1996,freeman-PRB-2009}. Incommensurate paramagnon spin excitations exist to well beyond the spin and charge ordering temperatures into a charge-stripe liquid phase.}
\label{stripeliquid}
\end{center} \end{figure}

 At half doping below $\sim$ 480\, K a stable two dimensional checkerboard charge ordered state is observed, where every other Ni site is part of the charge order and the charge order is along both diagonal directions of the Ni-O planes\cite{chen93,kajimoto-PRB-2003}. Checkerboard charge-order is a spin frustrated state, as the spins try to order as stripes but their stripe direction is undefined between two possible directions. On cooling below 180\,K the checkerboard charge order changes to part checkerboard charge order and part charge-stripe order, with only spin stripe order with a periodicity that matches the charge-stripe order occurring at lower temperature\cite{kajimoto-PRB-2003}. Studies of both the magnetic excitation spectrum by inelastic neutron scattering and magnetic order by $\mu$ Spin Rotation spectroscopy confirm that the spin stripe order occurs across the full material which has mixed charge order character\cite{freeman-PRB-2005,williams-PRB-2016}. In the magnetic structure of half doped LSNO the spin order will gain $J$ intra-spin stripe interactions by having neighbouring spins in spin stripes\cite{freeman-PRB-2005}. This results in a defined spin stripe direction that stabilizes the order within a domain.  It is likely that the spin fluctuations of a stripe liquid phase drive the change in charge order structure. In figure \ref{stripeliquid} we show how spin stripe correlations persist in charge-stripe ordered La$_ {2}$NiO$_{4.11}$ to temperatures well in excess of the spin ordering temperature of $122.5\pm2.5$\,K. This is consistent with the evidence for charge-stripe correlation occurring above the charge ordering temperature occurring in a nematic stripe liquid phase\cite{Stripeliquidnematic}. 
 
For charge-stripe order below one third doping we have noted that $\varepsilon$ is closer to one third than expected from the chemical doping level, so the average charge-stripe spacing and the average width of the spin stripes are both reduced. This results in a reduction in the number of the ordered Ni$^{2+}$ $S = 1$ moments and their intra-spin stripe interactions $J$, although with a well defined spin stripe direction. The energetic loss of stability to the magnetic order for $\varepsilon$ being greater than the chemical doping level has to be offset by the reduction of other interactions. At a third doping there is no need for self doping to obtain an ideal body centred tetragonal stacking, therefore the spin interactions are no longer in competition with minimisation of the out of plane Coulomb interaction.


\subsection{Spin and Charge Correlation lengths}

In LSNO there is a difference between the spin and charge ordering correlation lengths\cite{yoshizawa-PRB-2000,Lee-PRL-1997}. The correlation lengths for the spin order are larger than for the charge order, for the in plane correlation lengths parallel and perpendicular to the charge-stripe direction, and for the correlation lengths out of the Ni-planes. In the Ni-O planes of LSNO $x = 1/3$ the spin ordering correlation lengths are three times longer than that of the charge-stripe correlation length\cite{Lee-PRL-1997}. Naively one may expect that if the spin and charge-stripes order is in the same domain the two orders would have the same correlation length. Modelling examined the effect of disorder from topological defects such as ends of charge-stripes, or from elastic stripe deformations without topological defects, on the  spin- and charge-stripe correlation lengths within the plane\cite{Zachar}. Based on the principle that if the accumulated deviation of the order changes by half the periodicity of the order, the different regions within a domain are uncorrelated. It can be shown that if the in-plane spin correlation length is between 1-4 times longer than the charge-stripe correlation length, then the disorder is primarily from elastic deformation of the charge-stripes, i.e. the charge-stripes of LSNO. This is consistent with theoretical calculations that show electron hopping stabilizes charge-stripe order\cite{Poilblanc-PRB-1989,Raczkowski-PRB-2006}.

At one third doping the ability to flex within the plane  is constrained by the spatial extent of the Ni centred charge-stripes being predominantly on the O surrounding the Ni, which should effect the charge of the neighbouring in-plane magnetically ordered Ni$^{2+}$ sites\cite{schu-PRL-2005}. When a charge-stripe flexes onto the neighbouring Ni site it would physically be in contact with the neighbouring charge-stripe so that further flexing is likely to cause the charge-stripe to flex back to 3 Ni-Ni spacings apart. Thus at one third doping the charge-stripes gain rigidity, and the in plane correlation lengths have a maximum\cite{yoshizawa-PRB-2000}. With long range well correlated in-plane order, the charge-stripe order between planes will also be well correlated, and results in a shallow maximum of the out of plane correlation lengths\cite{yoshizawa-PRB-2000}. We note however that at lower doping levels, the correlation lengths in LSNO are longer when the hole doping is from excess O not Sr substitution, this is likely to be due to the order locking into an O supercell structure\cite{Tranquada-PRL-1994}.

As discussed in section 4 for incommensurately doped LSNO the spacing between the charge-stripes varies to achieve the average periodicity. Below one third doping, the charge-stripe order has locations where the charge-stripes are spaced more than 3 Ni-Ni spacings apart within the Ni-O plane, and there will be little energy cost if the charge-stripe spacing to the next charge-stripe is 4 Ni-Ni spacings to the next charge-stripe in front or behind. In this way the charge-stripes will have a greater freedom to flex within the Ni-O plane, and these flexes would be able to propagate in the ordered structure. This limits the correlation length within the plane as the charge-stripe flexing  decorrelates the different regions of a domain, resulting in the in plane correlation length decreasing on reducing doping away from one third\cite{yoshizawa-PRB-2000}. The charge-stripe flexing within the Ni-O plane will also cause the charge order between Ni-O planes to decorrelate.  To the best of the authors knowledge, there is no stabilization of charge-stripe order in LSNO at $n_h = 1/5$ doping or $\varepsilon = 1/5$.  At $n_h = 1/5$ the charge-stripe order can stack with an ideal body centred tetragonal structure to minimize the Coulomb repulsion between charge-stripes of different layers. However the flexing of the charge-stripes with a large charge-stripe spacing,  is less energetically constrained by the Coulomb interaction. Unlike at $1/3$ doping the charge-stripes at $1/5$ doping in LSNO gain little rigidity from obtaining a body centred tetragonal structure, and the correlation length of the charge-stripe order at $\varepsilon = 1/5$  is observed to be short ranged\cite{Sachan-PRB-1995}.

\section{Discussion and Conclusions}

In this paper we have discussed experimental and theoretical studies of LSNO to understand the stability of charge-stripe order at one third doping in this material. The key elements are the Coulomb interaction between charge-stripes that is strong out-of the Ni-O plane\cite{wochner-PRB-1998}, a site preference for the charge-stripes\cite{wochner-PRB-1998,li-PRB-2003,schu-PRL-2005}, with a small energy difference between O and Ni centring\cite{wochner-PRB-1998,li-PRB-2003}, the extended in plane nature of the holes\cite{li-PRB-2003,schu-PRL-2005}, electron hopping of the holes of the charge-stripes\cite{Poilblanc-PRB-1989,Zaaneen-PRB-1994,Raczkowski-PRB-2006}, one hole per charge-stripe lattice site\cite{Zaaneen-PRB-1989,Poilblanc-PRB-1989,Zaaneen-PRB-1994,Katsufuji-PRB-1999} and the availability of low energy conduction states\cite{Katsufuji-PRB-1999}. The out Ni-O plane Coulomb interaction for purely Ni centred charge-stripes favours a body centred structure of charge-stripes running parallel to each other in different planes, and an odd number of Ni-Ni spacings between charge-stripes. With preference for Ni centring of charge stripes in LSNO, charge-stripe order with an even number of Ni-Ni spacings does not minimize the out-of-plane Coulomb interaction. Below one third doping the availability of low lying conduction states allows for self doping so the hole doping available for charge-stripe order is nearer to one third doping. The electron hopping that aids charge-stripe ordering, and the small energy difference between Ni and O centred charge-stripes allows the charge-stripes to flex. The extended size of the holes restricts the flexing for one third doped LSNO. At one fifth doping there is however a smaller energy cost associated with flexing of the charge-stripes so that flexing charge-stripes don't form long range order, with an ideal body centred tetragonal structure that minimizes the out-of-plane Coulomb interaction. Above one third doping there is an increased cost of the  in- and out-of-plane Coulomb interaction with increasing doping, along with a reduction in the magnetic spin interactions, all of which  favour the charge-stripe order being closer to one third doping than for charge-stripes with one hole per charge site and the chemical doping level of the material.
 
The correlation length of charge order out of the Cu-O planes in the cuprates is short ranged\cite{tranquadaaxxe-PRB-1996}, implying that the inter-CuO plane interactions of the charge order are weaker than the in plane interactions. The out-of-plane Coulomb interactions between charge-stripes in LSNO leads to parallel charge-stripes in adjacent Ni-O layers and the structure of the charge-stripe order perpendicular to these planes\cite{wochner-PRB-1998}. We have argued in this paper how this is one of the key factors in LSNO that results in a charge-stripe order stability at one third doping. The balance of interactions in the cuprates that result in a stability of charge order at one eighth doping are likely to differ from the interactions in LSNO. We should however consider similarities between the cuprates and LSNO, such as the probable extended nature of holes implied from the covalent nature of bonding in cuprates\cite{Walters}, when trying to understand the origin of the one eighth doping stability in the cuprates. From the insights of our review into the one third doping stability of charge-stripe order in LSNO, we hope to spur on theoretical modelling of charge-stripe order in LSNO, and provoke further discussion of the one eighth charge order stability of the cuprates.

We would like to note that during the review process of this paper a phenomenological model of the incommensurability of charge-stripe order in LSNO has appeared\cite{Bucher}. This model predicts the
 incommensurability of LSNO well over the doping range $0.27 \leq n_{h} \leq 0.5$, and is based on charge-stripe order in LSNO occuring above a critical doping level of  $1/9\, th$.

\section{Acknowledgements}

This research project has been supported by the European Commission under the 7th Framework Programme through the 'Research Infrastructures' action of the `Capacities' Programme, Contract No: CP-CSA\_INFRA-2008-1.1.1 Number 226507-NMI3. This work is based on experiments performed at the Swiss spallation neutron source SINQ, Paul Scherrer Institute, Villigen Switzerland, the Institut Laue-Langevin, Grenoble France, and the Heinz Maier Leibnitz Zentrum (MLZ), Garching, Germany. Crystal growth was supported by the Engineering and Physical Sciences Research Council of Great Britain.

\end{document}